\documentclass[reprint,aps,pra,floatfix]{revtex4-1}

\usepackage{color}

\usepackage{graphicx}
\usepackage{bm}

\begin{document}

\title{Structural and magnetic properties of GdCo$_{5-x}$Ni$_x$}

\author{Amy L. Tedstone}
\author{Christopher E. Patrick}
\author{Santosh Kumar}
\author{Rachel S. Edwards}
\author{Martin R. Lees}
\author{Geetha Balakrishnan}
\author{Julie B. Staunton}
\email{J.B.Staunton@warwick.ac.uk}
\affiliation{Department of Physics, University of Warwick, Coventry, CV4 7AL, United Kingdom}
\date{\today}

\begin{abstract}
GdCo$_5$ may be considered as two sublattices---one of Gd and one of Co---whose 
magnetizations are in antiparallel alignment, forming a ferrimagnet. 
Substitution of nickel in the cobalt sublattice of GdCo$_5$ has been 
investigated to gain insight into how the magnetic properties of this 
prototype rare-earth/transition-metal magnet are affected by changes in 
the transition metal sublattice.
Polycrystalline samples of GdCo$_{5-x}$Ni$_x$ for 0~$\leq x \leq$~5 
were synthesized by arc melting. 
Structural characterization was carried out by powder x-ray diffraction 
and optical and scanning electron microscope imaging of metallographic slides, 
the latter revealing a low concentration of Gd$_2$(Co, Ni)$_7$ lamellae for $x \leq 2.5$. 
Compensation---i.e.\ the cancellation of the opposing Gd and transition metal moments--- 
is observed for $1 < x < 3$ at a temperature which increases with Ni content;
for larger $x$, no compensation is observed below 360~K.
A peak in the coercivity is seen at $x \approx 1$ at 10~K 
coinciding with a minimum in the saturation magnetization.
Density-functional theory calculations within the disordered local 
moment picture reproduce the dependence of the magnetization on 
Ni content and temperature.
The calculations also show a peak in the magnetocrystalline anisotropy 
at similar Ni concentrations to the experimentally-observed coercivity maximum.
\end{abstract}


\maketitle

\section{Introduction \label{sec.intro}}

\begin{figure}[b]
\includegraphics{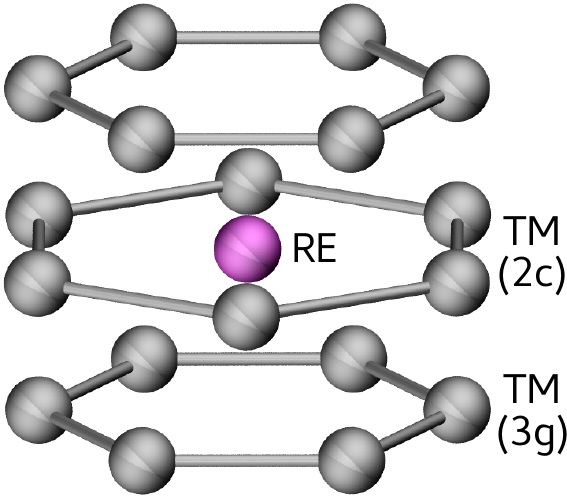}
\caption {Ball-and-stick model of the CaCu$_5$ structure adopted by RETM$_5$ compounds,
showing the rare earth site (purple) and inequivalent 2$c$ and 3$g$ 
transition metal sites (gray).}
\label{fig.ball_stick}
\end{figure}

The RETM$_5$ [(RE = rare earth; TM = transition metal)] family of materials have widely 
ranging magnetic properties owing to the differing number of 4$f$ electrons found 
in the RE elements.
These materials crystallize into a hexagonal lattice (the CaCu$_5$ type structure, 
space group $P6/mmm$, Fig.~\ref{fig.ball_stick}) with a unit cell consisting of layers with a central RE atom 
surrounded by TM atoms in $2c$ positions, alternating with layers of TM atoms 
in the $3g$ positions~\cite{Kumar1988}. 
One pertinent example is SmCo$_5$, a permanent magnet which can be favored over 
Nd-Fe-B magnets for its superior high-temperature performance (Curie temperature 
of around 1020~K~\cite{Strnat1967}, as opposed to approximately 580~K for 
Nd-Fe-B magnets~\cite{Sagawa1984,Croat1984}). 
Another member of this family is GdCo$_5$, where the symmetry of the Gd $4f$ 
shell causes crystal-field effects to vanish~\cite{Kuzmin2008}.
The absence of crystal-field effects make GdCo$_5$ a particularly useful system 
to study the rare-earth/transition-metal interaction via both theory and experiment. 

GdCo$_5$ is ferrimagnetic.
Starting from $T=0$~K its magnetization increases with increasing temperature 
reaching a maximum at around 800~K~\cite{Nesbitt1959, Nassau1960}.
With a further increase in temperature, the spontaneous magnetization decreases 
to zero at the Curie temperature (1014~K)~\cite{Buschow1977}.
This unusual temperature dependence is a consequence of the Gd moments disordering 
more rapidly with temperature than the Co moments ~\cite{Yermolenko1980,Patrick2017}.

Doping a RETM$_5$ material can change its magnetic properties in a controlled 
manner~\cite{Patrick2017, Bajorek2011, Grechishkin2006, deOliveira2011, 
Buschow1976, Chuang1982, Chuang1981, Liu1991, Burzo1999, Drzazga1989}. 
Here, only doping of the TM sublattice is considered. 
The effects of doping on coercive field and saturation magnetization 
have been studied for single crystals of GdCo$_{5-x}$Cu$_x$ by 
Grechishkin~\textit{et al}.~\cite{Grechishkin2006} and later 
by de Oliveira~\textit{et al}.~\cite{deOliveira2011}.
The Cu reduces the TM sublattice magnetization.
Both papers report a peak in coercivity at a composition of $x \approx 1.5$.
This is found to be the compensation composition of this intermetallic 
at room temperature, where the (Co,Cu) sublattice magnetization 
exactly cancels (fully compensates) the Gd sublattice magnetization.
A peak in coercivity was also found in YCo$_{5-x}$Ni$_x$~\cite{Buschow1976} 
and for RECo$_{5-x}$Ni$_x$ (RE = Sm, La, Y, Th and Ce)~\cite{Chuang1982} 
for certain compositions.
It is interesting that RECo$_5$ compounds containing nonmagnetic REs 
such as Y and La still exhibit a peak in coercivity, despite not having a compensation composition.
Buschow and Brouha (Ref.~\citenum{Buschow1976}) suggested that the presence of narrow 
Bloch walls in YCo$_{5-x}$Ni$_x$ is primarily responsible for the high 
coercivity observed for certain compositions.

In a previous work~\cite{Patrick2017} we prepared polycrystalline samples 
of GdCo$_{5-x}$TM$_x$ and YCo$_{5-x}$TM$_x$ (TM = Ni and Fe) with $x \leq 1$ 
and single crystal GdCo$_5$ and YCo$_5$, and compared the experimentally 
determined magnetic properties of these samples with theoretical 
calculations made using density-functional theory.
An increase (decrease) in magnetization was observed for Fe (Ni) doping.
The calculations showed that substituting Ni onto the Co lattice led to 
Ni preferentially occupying the $2c$ site,
although experimentally this may depend on the method of sample preparation.
The doping site was not found to have a large effect on magnetization, 
but did affect the Curie temperature, with a larger change for the $2c$ site doping.
However, possible effects on the coercivity and magnetocrystalline anisotropy 
were not explored in the paper, nor were concentrations with $x > 1$.

The magnitude of the Ni moment in GdCo$_{5-x}$Ni$_{x}$ is very small~\cite{Bajorek2011} 
compared to that of Co and Gd (in GdCo${_5}$, the Co moment 
is $\approx~1.6\mu_{\mathrm{B}}$/atom at both the $2c$ and the $3g$ sites 
and the Gd moment is $\approx~7\mu_{\mathrm{B}}$~\cite{Patrick2017}), 
hence the (Co, Ni) sublattice magnetization is expected to decrease 
with increasing Ni content.
The fully substituted material GdNi$_5$ is ferrimagnetic, but the main 
contribution to the magnetization comes from the ferromagnetic Gd 
sublattice, giving a Curie temperature of 32~K~\cite{Gignoux1976, Bajorek2011}.
At absolute zero at a particular composition for GdCo$_{5-x}$Ni$_x$ the 
(Co, Ni) sublattice magnetization will fully compensate the Gd 
sublattice magnetization.
The compensation composition at absolute zero will fulfill the 
condition $\mu_{\mathrm{Gd}}~-~\mu_{\mathrm{Co}}(5-x)-\mu_{\mathrm{Ni}}x~=~0$.
Taking approximate zero temperature values for the moments of Co, Ni, and 
Gd (1.6,~0.6,~7$\mu_{\mathrm{B}}$/f.u.\ [formula unit]), respectively, 
the compensation composition is $x\sim1$.
At other compositions there may exist a finite compensation temperature 
where the different disordering of the Gd and (Co, Ni)
sublattice magnetizations again leads to compensation.

Chuang~\textit{et al}.~\cite{Chuang1981} replaced Co with Ni in 
GdCo$_{5-x}$Ni$_x$ for $0\leq x\leq 5$ measuring magnetization 
versus temperature at 12~kOe for several compositions, 
focusing primarily on temperatures from 300 to 1015~K, with the 
exception of GdCo$_2$Ni$_3$ and GdCoNi$_4$ for which measurements 
were taken over the ranges of 77--1015~K and 77--300~K, respectively.
Therefore, for compositions with $x<3$, any compensation point 
at temperatures lower than 300~K would not have been observed.
GdCo$_{5-x}$Ni$_x$ has been investigated by Liu~\textit{et al}.~\cite{Liu1991} 
for $x \leq 1.05$ in order to determine the intersublattice RE/TM coupling constant.
Magnetization compensation has also been studied in Gd(Co$_{4-x}$Ni$_x$)Al~\cite{Burzo1999}, 
where increased Ni content was observed to increase the compensation 
temperature and in RE(Co$_{4-x}$Fe$_x$B) (RE = Gd and Dy)~\cite{Drzazga1989}, 
where the compensation temperature was reduced for increased Fe content.

In this paper the magnetic behavior of polycrystalline powders and buttons of 
GdCo$_{5-x}$Ni$_x$ is reported for temperatures from 5 to 360~K for 
$x=0$, 0.5, 1, 1.25, 1.28, 1.3, 1.5, 2, 2.5, 3, 3.5, 4, and 5. 
The compensation temperature, coercivity and magnetization at 70~kOe are 
presented as a function of $x$.
This extends the previous work of Chuang~\textit{et al}.~\cite{Chuang1981} 
to a temperature range in which the compensation point can be observed for $x<3$.
The behavior is then analyzed with the help of density functional theory 
calculations within the disordered local moment picture~\cite{Gyorffy1985, Staunton2006}, 
calculating the composition-dependent magnetization, coercivity, and 
magnetocrystalline anisotropy.
The calculations provide microscopic insight into the macroscopic quantities 
observed experimentally, demonstrating the utility of the joint 
computational/experimental approach in understanding the 
behavior of RE/TM permanent magnets.

The rest of this paper is organized as follows.
Section~\ref{methodology} describes the experimental and theoretical
techniques used.
Sections~\ref{results} and \ref{sec.theory} describes the results
of the experiments and calculations, respectively.
The conclusions and summary are presented in Section~\ref{summary}.

\section{Methodology}\label{methodology}

\subsection{Experimental approach}

Polycrystalline samples of the series GdCo$_{5-x}$Ni$_x$ were synthesized by arc 
melting the constituent elements on a water cooled copper hearth under an argon atmosphere. 
The starting elements (99\% purity) were taken in the stoichiometric ratios with 1\% 
excess of Gd to compensate for losses during melting.
To ensure homogeneity, the ingots were flipped and remelted at least three times.
Annealing for 10 days at 950 $^\circ$C was tried to improve phase homogeneity, 
as discussed by Buschow and den Broeder~\cite{Buschow1973}.
However, analysis of the annealed samples (via the same methods described below) 
showed no convincing evidence that annealing promotes the 1:5 [Gd:(Co, Ni)] 
phase formation over the neighboring phases of 2:17 and 2:7, and so as-cast 
samples were used (provided they were found to be sufficiently phase pure, 
as described in the remainder of this section). 

The structures were characterized by powder x-ray diffraction using a 
Panalytical Empryean diffractometer with Co$K_\alpha$ radiation.
To confirm the phase content, metallographic slides were prepared from 
slices of ingots mounted in Epomet-F plastic and polished using 
progressively finer diamond suspensions.
Optical microscopy and scanning electron microscopy (SEM) 
imaging of the slides were used to further examine the structure.

Magnetization measurements were made as a function of temperature and 
applied field using a Quantum Design Magnetic Property Measurement 
System (MPMS) superconducting quantum interference device (SQUID) magnetometer.
Free to rotate powders were used to obtain the best estimate of saturation 
magnetization. 
The magnetization vs temperature, $M\left(T\right)$, data were taken in a 
10~kOe field with the temperature decreasing from 360 to 10~K at a 
rate of 3~K/min.
The magnetization versus field [$\left(M\left(H\right)\right)$] data 
were taken at 10~K (5~K for $x = 0$, 0.5, and 1) by first 
applying a 70~kOe field then collecting the magnetization data at 
incrementally decreasing fields (until 0~kOe).
Polycrystalline buttons were fixed to a sample holder using GE 
varnish to measure coercivity in an Oxford Instruments vibrating 
sample magnetometer.
The buttons were fixed in this manner to prevent sample rotation, 
and thus get a more accurate measurement of coercivity.
The coercivities were determined from four quadrant hysteresis loops 
starting at 70~kOe, as the coercivity is known to depend on the 
initial applied field~\cite{Katayama1976}.

\subsection{Theoretical approach}

The magnetic properties of Ni-doped GdCo$_5$ were calculated at 
zero and finite temperature using density-functional theory 
within the disordered local moment (DFT-DLM) picture~\cite{Gyorffy1985}.
In this approach, both the temperature-induced local moment 
disorder and the compositional disorder from the Ni doping are modeled
using the coherent potential approximation (CPA) within the Korringa-Kohn-Rostocker
(KKR) multiple-scattering formulation of DFT~\cite{Ebert2011}.
A detailed description of this approach applied to Ni-doped GdCo$_5$ 
is given in Ref.~\citenum{Patrick2017}; here the computational 
details specific to this work are given.

The calculations were performed on the CaCu$_5$ structure with 
lattice parameters fixed at $a= 4.979$~\AA, $c=3.972$~\AA, which 
were measured for pristine GdCo$_5$ at 300~K~\cite{Andreev1991}.
As discussed in Section~\ref{sec.theory} the Ni dopants were set 
to either occupy the 2$c$ and 3$g$ crystal sites with equal 
probability, or to preferentially sit at the 2$c$ crystal sites.
The KKR multiple-scattering equations were solved within the 
atomic sphere approximation (ASA) with Wigner-Seitz radii 
of (1.58, 1.39, 1.42) \AA \ at the (Gd, 2$c$, 3$g$) sites.

The KKR-CPA code \texttt{Hutsepot}~\cite{Daene2009} was used 
to generate scalar-relativistic potentials for the 
magnetically-ordered (ferrimagnetic) state,
expanding the key quantities in an angular momentum basis up 
to a maximum quantum number $l=3$. 
Exchange and correlation were treated within the local spin-density 
approximation~\cite{Vosko1980}, with the local self-interaction 
correction~\cite{Lueders2005} also applied to the Gd-$4f$ electrons.

The scalar-relativistic potentials were then fed into our own code which solves
the fully-relativistic scattering problem in the presence of magnetic 
disorder~\cite{Staunton2006}.
For selected Ni concentrations an orbital polarization correction 
(OPC)~\cite{Eriksson19901, Steinbeck2001} was included on 
the $d$ scattering channels~\cite{Ebert1996, PatrickStaunton2018}.
The DFT-DLM Weiss fields, which govern the temperature dependence 
of the calculated quantities, were calculated self-consistently 
using an iterative procedure~\cite{Matsumoto2014, Patrick2017}.
The magnetocrystalline anisotropy constants were calculated using 
the torque formalism described in Ref.~\citenum{Staunton2006}, using an 
adaptive reciprocal-space sampling scheme to ensure numerical accuracy~\cite{Bruno1997}.

To calculate magnetization versus field curves the first-principles approach 
to calculate temperature-dependent magnetization vs field (FPMVB) curves introduced
in Ref.~\citenum{Patrick2018} was used.
A set of 28 DFT-DLM calculations are used to fit the parameters contained 
in $F_2$, which quantify exchange and magnetic anisotropy.
The magnetization for a given field and sample orientation is then determined 
by minimizing the free energy with respect to the angles between the Gd 
and the transition metal magnetizations and the crystal axes.
The approach does not account for any canting between moments within the 
transition metal sublattices, since this was previously calculated to 
be less than 0.1$^\circ$ for GdCo$_5$~\cite{Patrick2018}.

\section{Results\label{results}}

\subsection {Structural characterization}\label{Structure}

\begin{figure}[tb]
\includegraphics[width=90mm]{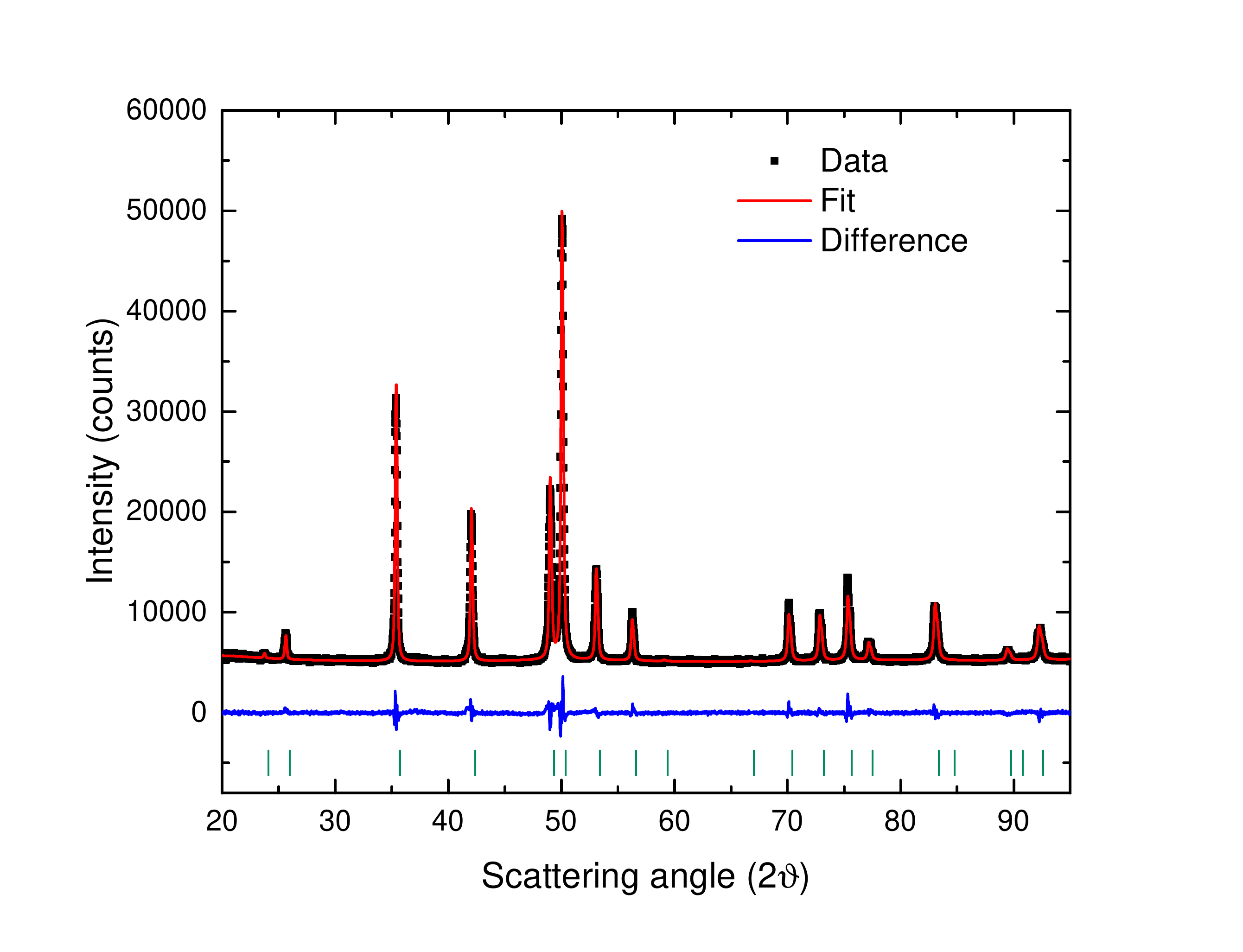}
\caption{\label{fig.refinement}The data and refinement (black 
data and red line respectively) for the powder x-ray diffraction 
carried out on GdCo$_3$Ni$_2$. 
The blue line shows the difference plot between the 
original data and the refinement.
The green ticks indicate indexed peaks.
The goodness of fit parameter is 2.03.}
\end{figure}

Figure \ref{fig.refinement} shows the powder x-ray diffraction 
pattern obtained from GdCo$_3$Ni$_2$ ($x=2$, black data). 
The red line is the fit obtained when a Rietveld refinement 
is carried out using the TOPAS software~\cite{Topas}.
The goodness of fit parameter is 2.03 (a similar value was 
found from fitting the diffraction patterns of all samples). 
The blue line is the difference plot between the observed data 
and the fit.
The green ticks represent indexed peaks.
All the peaks in each diffraction pattern have been indexed 
using the GdCo$_5$ structure---space group $P6/mmm$---as demonstrated here, 
suggesting the samples form as single phase materials to 
within the detection limits of the technique.
The lattice parameters for all $x$ are given in Table~\ref{tab.lp}, along 
with the lattice parameters found in the literature.
As expected from previous research, there is a contraction in 
the $ab$ plane with increasing Ni content, 
and a less pronounced general contraction along the $c$ axis.

\begin{table}[tb]
\begin{ruledtabular}
\caption{\label{tab.lp}Lattice parameters of GdCo$_{5-x}$Ni$_x$ 
obtained from Rietveld refinement of the powder x-ray 
diffraction patterns.
The results reported previously in literature are included for comparison. }
\begin{tabular}{ccccc}

{} & {} & {} & Reported & Reported \\
{$x$} & { $a$ ({\AA})} & {$c$ ({\AA})} & $a$~({\AA}) & $c$~({\AA}) \\
\hline
$0$ & 4.946(9) & 3.999(7) & 4.979\footnotemark[1] & 3.972\footnotemark[1] \\
{} & {} & {} & 4.960\footnotemark[2] & 3.989\footnotemark[2] \\
{} & {} & {} & 4.974\footnotemark[3] & 3.973\footnotemark[3]\\ 
$0.5$ & 4.9680(4) & 3.9790(3) & - & -\\
$1$ & 4.9681(4) & 3.9794(3) & 4.959\footnotemark[1] & 3.977\footnotemark[1]\\
$1.5$ & 4.957(1) & 3.9790(5) & - & - \\
2 & 4.9493(1) & 3.9803(1) & 4.948\footnotemark[1] & 3.980\footnotemark[1]\\
2.5 & 4.9439(1) & 3.9785(1) & 4.94\footnotemark[1] & 3.979\footnotemark[1]\\ 
3 & 4.9338(2) & 3.9738(2) &  4.932\footnotemark[1] & 3.967\footnotemark[1]\\
3.5 & 4.9306(1) & 3.9708(1) & - & - \\
4 & 4.9245(1) & 3.9704(1) & 4.92\footnotemark[1] &  3.969\footnotemark[1] \\
5 & 4.9139(1) & 3.9683(1) & 4.909\footnotemark[1] & 3.965\footnotemark[1] \\
{}& {} & {} & 4.91\footnotemark[4] & 3.967\footnotemark[4]\\ {} & {} & {}&4.90\footnotemark[5]& 3.97\footnotemark[5] \\
\footnotetext[1]{Ref.~\citenum{Chuang1981}}
\footnotetext[2]{Ref.~\citenum{Lihl1969}}
\footnotetext[3]{Ref.~\citenum{Wernick1959}}
\footnotetext[4]{Ref.~\citenum{sorgic1995}}
\footnotetext[5]{Ref.~\citenum{lemaire1969}}
\end{tabular}

\end{ruledtabular}
\end{table}

\begin{figure}[tb]
\includegraphics[width=70mm]{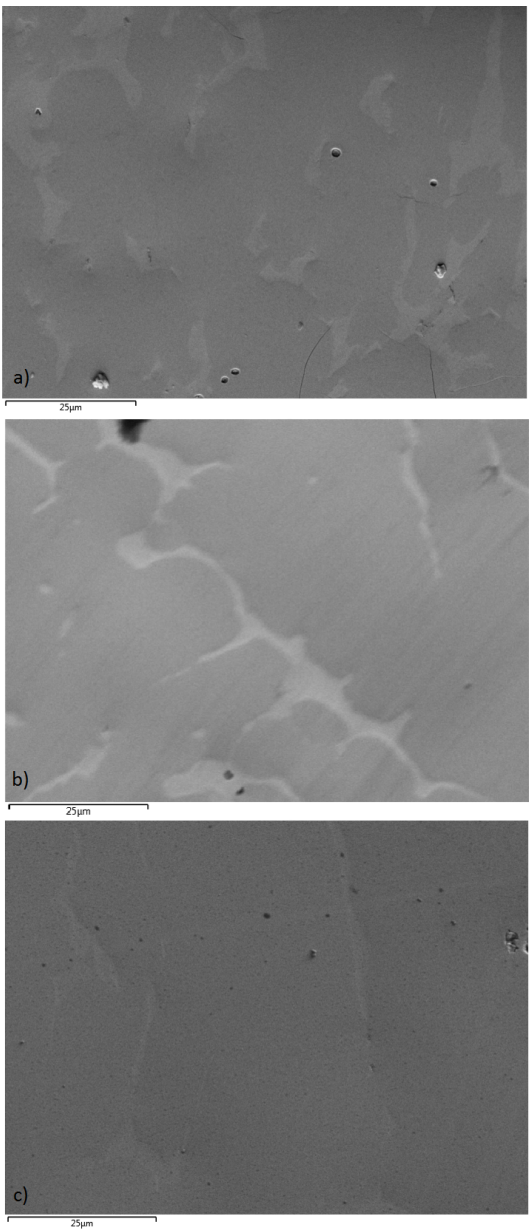}
\caption{\label{fig.SEM} SEM images  (secondary
electron mode) of 
(a) $x~=~0$, (b) $x~=~1.5$, and (c) $x~=~2$.
The secondary 2:7 phase can clearly be seen as 
lamellae which are lighter gray than the surrounding 
1:5 matrix.
Fig.~\ref{fig.SEM}(c) is relatively clear of lamellae.}
\end{figure}

The x-ray diffraction measurements show that the samples are single phase, 
but this technique may miss small percentages of impurity phases.
For this reason, optical microscopy and SEM images were taken of 
metallographic slides; some example SEM images are shown in 
Fig.~\ref{fig.SEM}.
A decreasing quantity of lamellae of a secondary phase were 
observed with increasing $x$, until $x~=$~2.5, beyond which 
no lamellae were observed.
Energy-dispersive x-ray spectroscopy showed that the majority 
phase is the 1:5 [Gd:(Co, Ni)] phase and the small lamellae 
are a 2:7 phase.
Table~\ref{tab.secondphase} shows the percentage of the 2:7 phase 
found in the as-cast samples.
Buschow and den Broeder noted that a slight excess of Co 
during the arc-melting promotes the formation of the 
2:7 phase within the 1:5 matrix~\cite{Buschow1973}.
GdNi$_5$ forms congruently from the melt, and is stable down 
to, at least, room temperatures.
On the other hand, GdCo$_5$ undergoes eutectic decomposition 
at 775~$^\circ$C into Gd$_2$Co$_7$ (2:7 Gd:Co) and 
Gd$_2$Co$_{17}$ (2:17 Gd:Co) \cite{Buschow1977} and so 
increasing Ni content improves the stability of the 1:5 phase.

\begin{table}[]
\begin{ruledtabular}
\centering
\caption{
\label{tab.secondphase}
Percentage of the 2:7 phase in GdCo$_{5-x}$Ni$_x$ 
for $x=1$, 1.25, 1.5, and 2 determined from 
optical and SEM images. }
\begin{tabular}{cc}
{$x$} & {2:7 phase (\%)} \\
\hline
1 & 5.2 \\
1.25 & 6.7 \\
1.5 & 5.7 \\
2 & 2.1\\
\end{tabular}

\end{ruledtabular}
\end{table}

The relative unimportance of the secondary 2:7 phase in 
these quantities (2--7\%) to the measurement of intrinsic 
quantities such as the magnetization can be demonstrated 
for the case of no Ni doping.
The moment per formula unit of GdCo$_5$ is 1.37$\mu_{\mathrm{B}}$/f.u.~\cite{Buschow1977} 
and of Gd$_2$Co$_7$ is 2.5$\mu_{\mathrm{B}}$/f.u.~\cite{Buschow1977}.
Assuming no other impurities or other phases, taking 93\% of the total 
powder to be GdCo$_5$ and 7\% to be Gd$_2$Co$_7$ (i.e. the maximum 
amount of the impurity phase estimated) the total moment becomes 
1.45$\mu_{\mathrm{B}}$/f.u.
This is an $\sim$6\% increase to pure GdCo$_5$; which is 
insignificant compared to the almost 300\% increase in moment 
from GdCo$_5$ to GdNi$_5$.
On the other hand, it is possible that the presence of a 
secondary phase affects extrinsic properties such as the 
coercivity, as in Sm-Co or Nd-Fe-B magnets~\cite{SepehriAmin2017}.
As no 2:7 phase was observed in as-cast samples of 
$x \geq 3$, these can be used as a comparison when 
studying trends in the measurements.

\subsection{Compensation temperature}\label{sec.Tc}

\begin{figure}[tb]
\includegraphics[width=90mm]{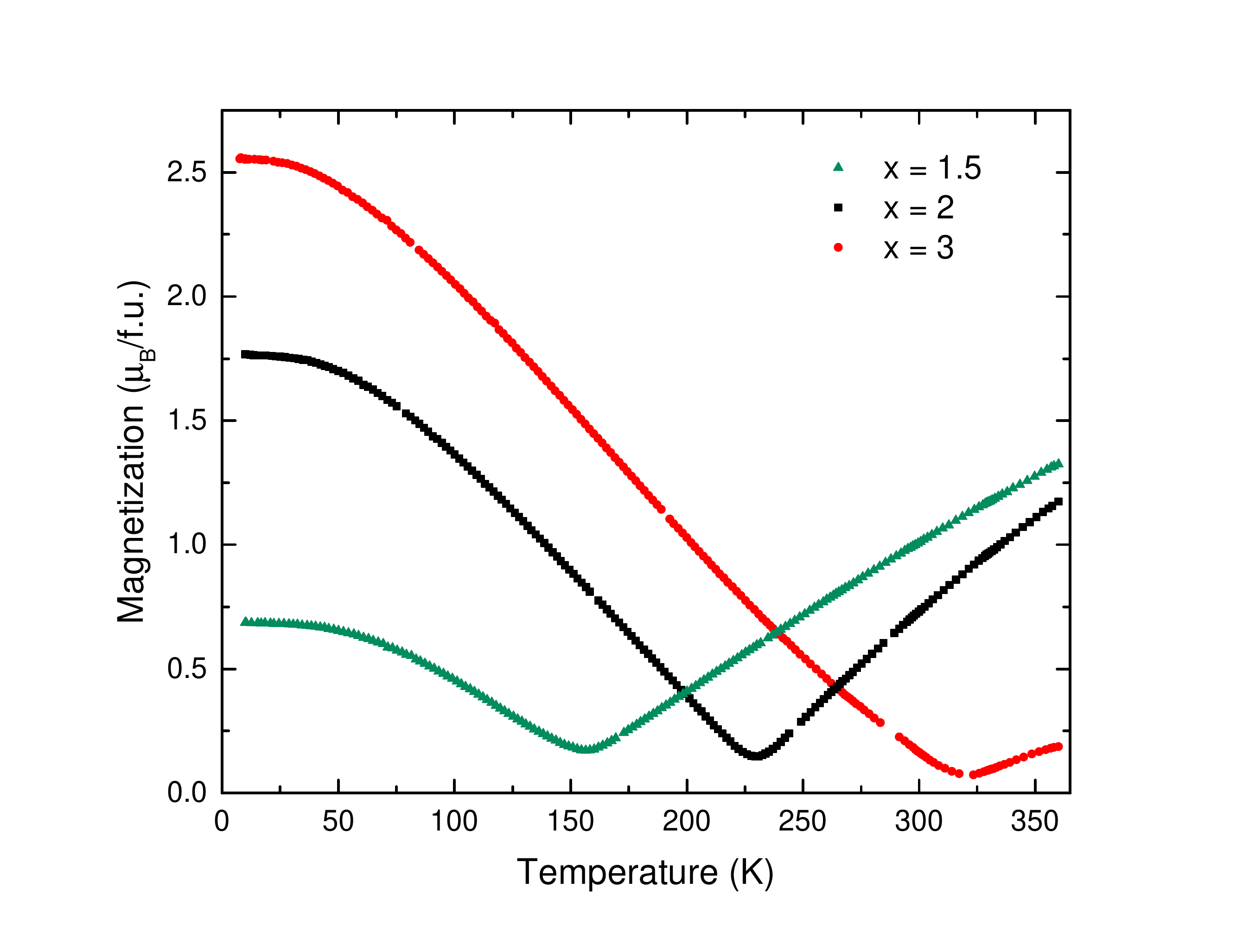}
\caption{\label{fig.MT} Magnetization of samples $x~=~1.5$ 
(green triangles); $x~=~2$ (black squares) and 
$x~=~3$ (red circles) versus temperature in a 10~kOe field.
A clear minimum can be seen at 157, 230, and 323~K respectively. }
\end{figure}

\begin{figure}[tb]
\includegraphics[width=90mm]{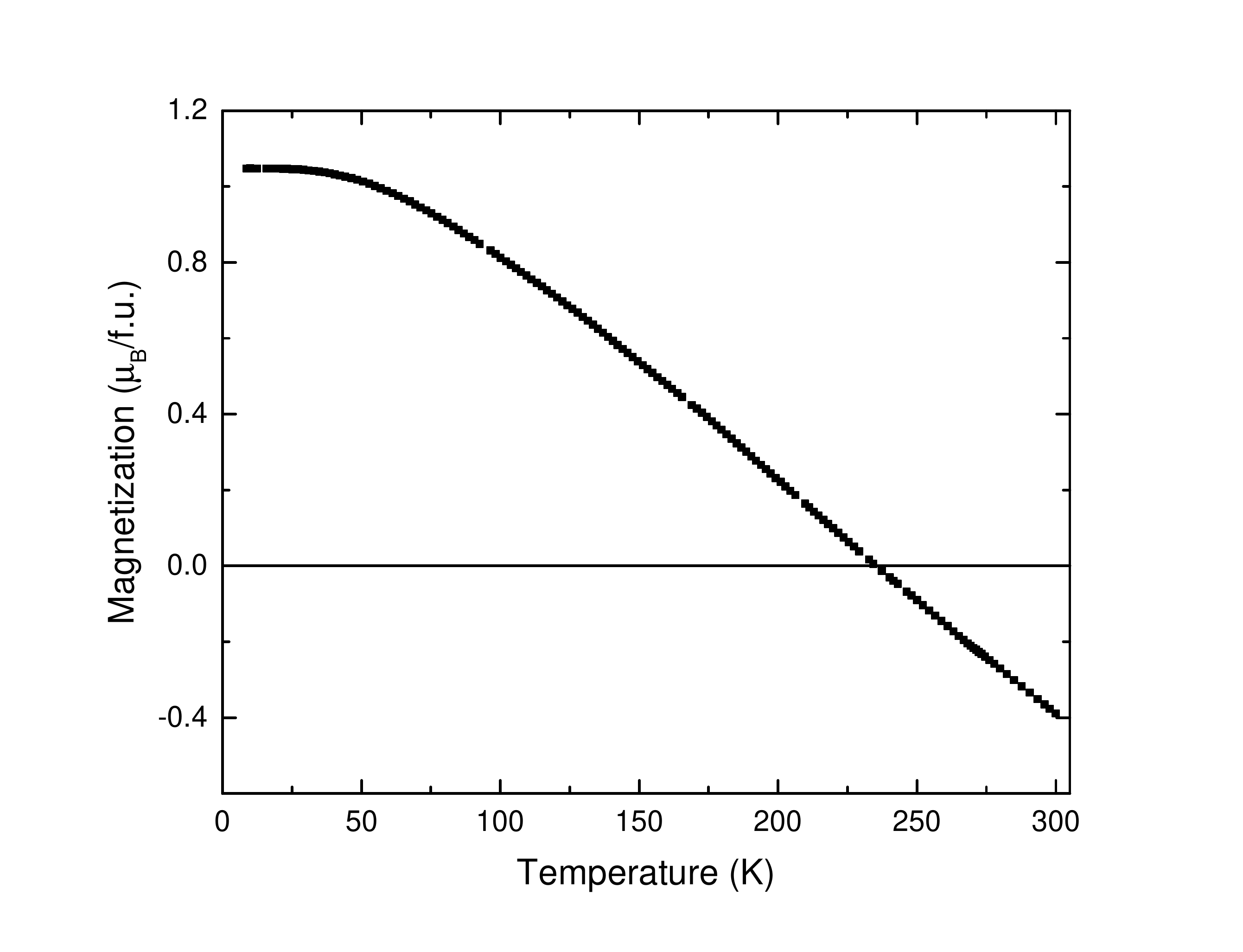}
\caption{\label{fig.MT0T} Magnetization versus temperature curve for $x~=~2$ 
in zero applied field with the temperature increasing from 
10 to 360~K after field cooling in a field of 10~kOe.
The magnetization changes sign at 235~K. }
\end{figure}

$M$ versus $T$ curves are shown in Fig.~\ref{fig.MT} for $x=1.5$, 
2, and 3 measured in an applied magnetic field of 10~kOe.
A clear minimum occurs at progressively lower temperatures 
as Ni content is decreased.
At this minimum the net total magnetization aligns with 
the bias field of 10~kOe.
In contrast, when the sample is warmed in the (small, negative) 
trapped field of the magnetometer magnet, after cooling 
in 10~kOe, the magnetization changes sign at the compensation 
temperature, as shown in Fig.~\ref{fig.MT0T} for $x=2$.

As shown in Fig.~\ref{fig.MT}, the magnetization of the powders
subject  to the 10~kOe applied field does not go to zero, even
at the compensation temperature.
An explanation for this behavior in terms of a change in the internal
magnetic structure of the powder particles (from antiparallel to canted RE/TM moments)
is given in Sec.~\ref{sec.twosub}.

\begin{figure}[tb]
\includegraphics[width=90mm]{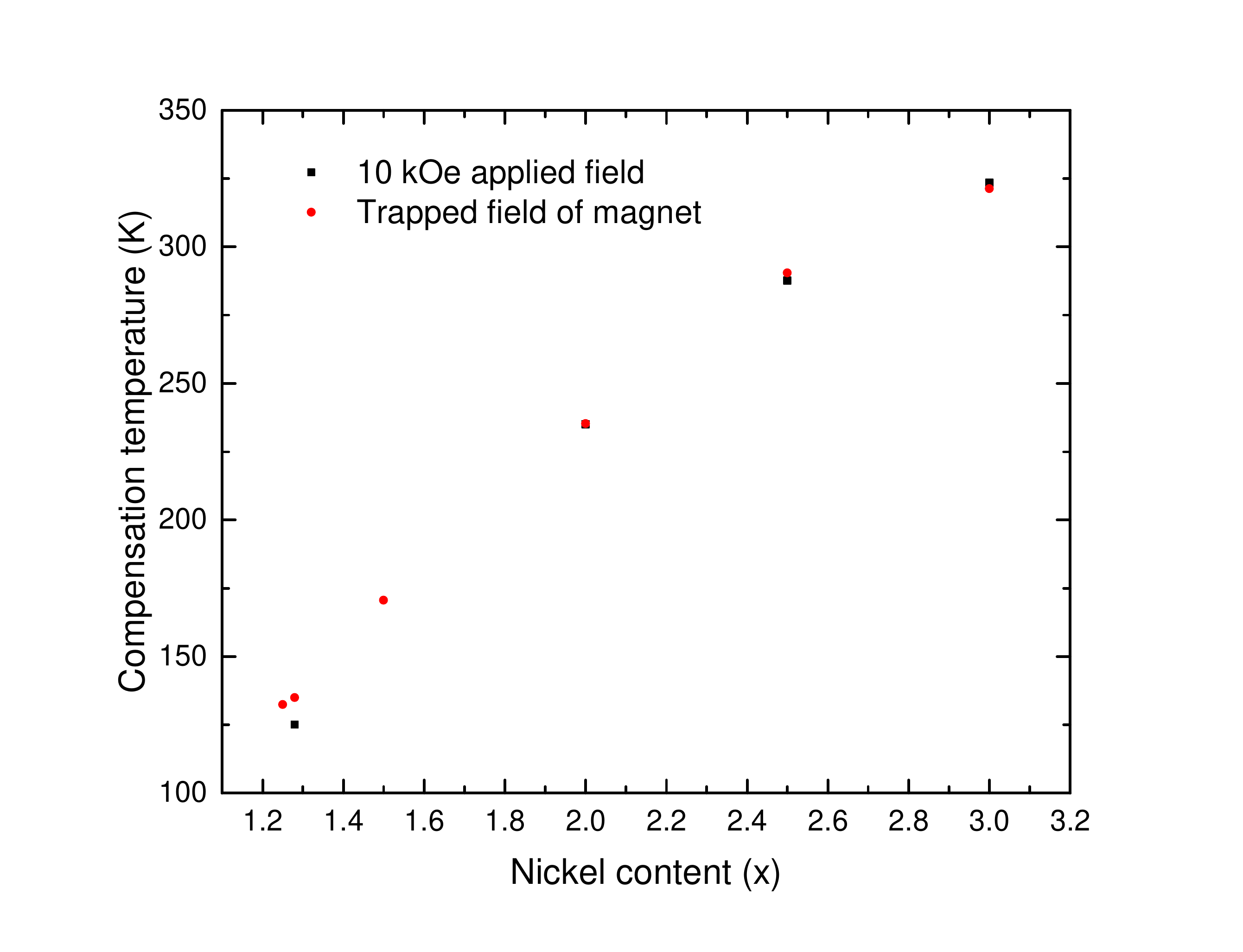}
\caption{\label{fig.Tcomp1T}Temperature of minimum magnetization 
(compensation temperature) in a 10~kOe field (black squares), 
and zero field (red circles) as a function of Ni concentration. }
\end{figure}

Fig.~\ref{fig.Tcomp1T} shows the dependence of the compensation 
temperature with composition.
A small hysteresis in compensation temperature with increasing/decreasing 
temperature was noted in a 10~kOe field for $x \leq 2$.
No hysteresis was found for these compositions in the trapped 
field of the magnet.
The compensation temperature increases for increasing nickel content, 
as expected.
Compensation is not observed for $x \geq 3.5$ and $x \leq 1$ in this 
temperature range.
In the case of low Ni doping, the Co sublattice magnetization, 
whilst reduced by the addition of Ni, remains dominant over 
the Gd sublattice for all measured temperatures.
At high Ni doping, the Co sublattice magnetization is reduced 
so much that the Gd sublattice magnetization dominates 
for all measured temperatures.

Chuang~\textit{et al}. (Ref.~\citenum{Chuang1981}) reported a 
compensation temperature for $x=3$ of 380~K in a 12~kOe field, 
however here it is 323~K in a 10~kOe field.
Earlier results used a large temperature step size, and 
hence some disagreement is to be expected.
Measurements here were obtained on as-cast powder 
samples, whereas Ref.~\citenum{Chuang1981} reports on annealed samples.

\subsection{Magnetization and coercivity}

\begin{figure}[tb]
\includegraphics[width=90mm]{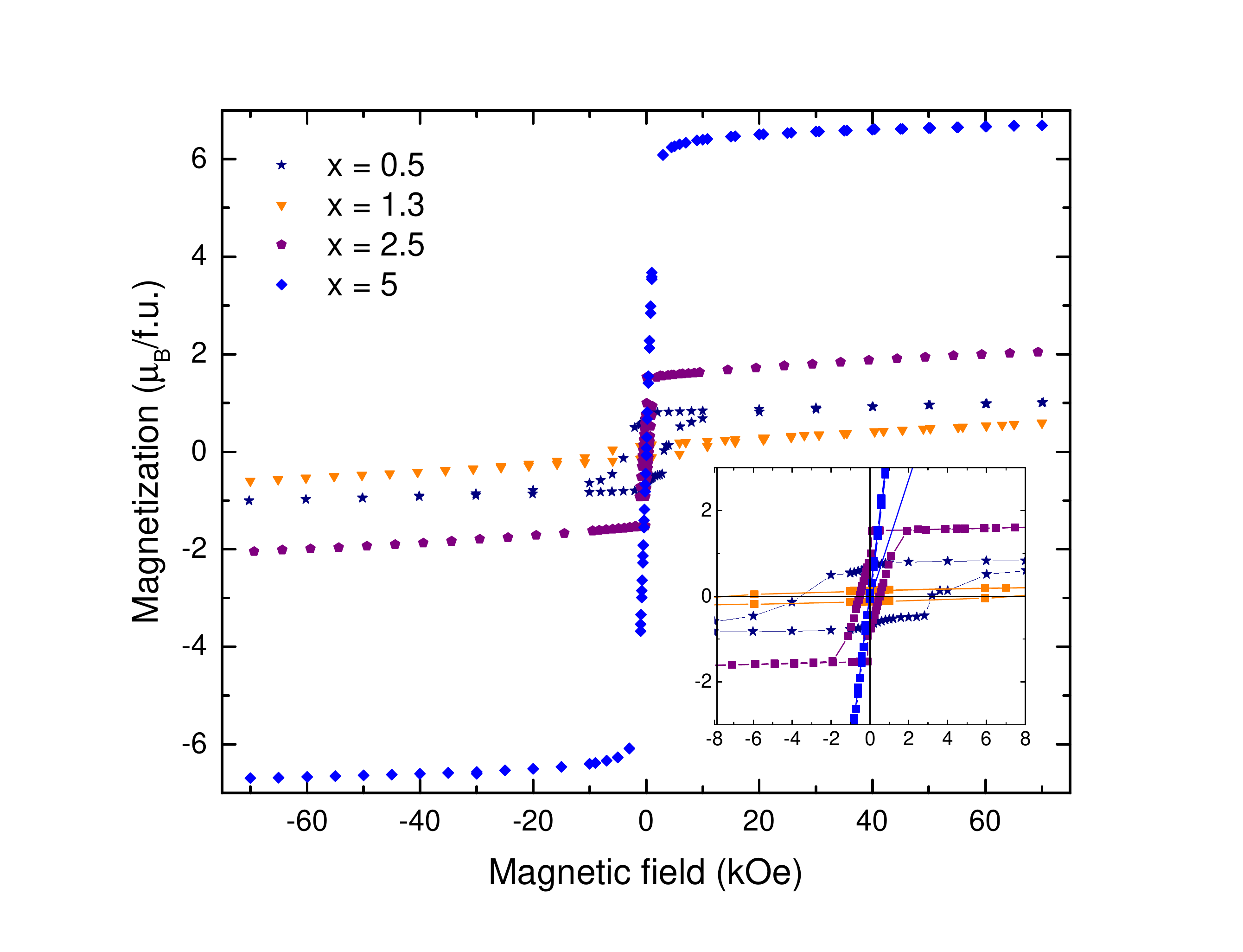}
\caption{\label{fig.MHloops} Four-quadrant $M$ versus $H$ loops of buttons 
of GdCo$_{5-x}$Ni$_x$ for $x=0.5$, 1.3, 2.5, and 5,  measured
at 10~K (5~K for $x$ = 0.5).
The inset shows the low field region of these $M$ versus $H$ loops.}
\end{figure}

Four-quadrant $M$ versus $H$ loops for buttons of composition 
$x=0.5$, 1.3, 2.5, and 5 are shown in Fig.~\ref{fig.MHloops}, 
with the inset showing the low-field region.
It is clear that the samples do not reach full saturation at 
70~kOe, which is the experimental limit.
This is to be expected, since the samples consist of a 
number of grains with randomly oriented $c$ axes, such that both 
easy and hard axes are being probed.

\begin{figure}[tb]
\includegraphics[width=90mm]{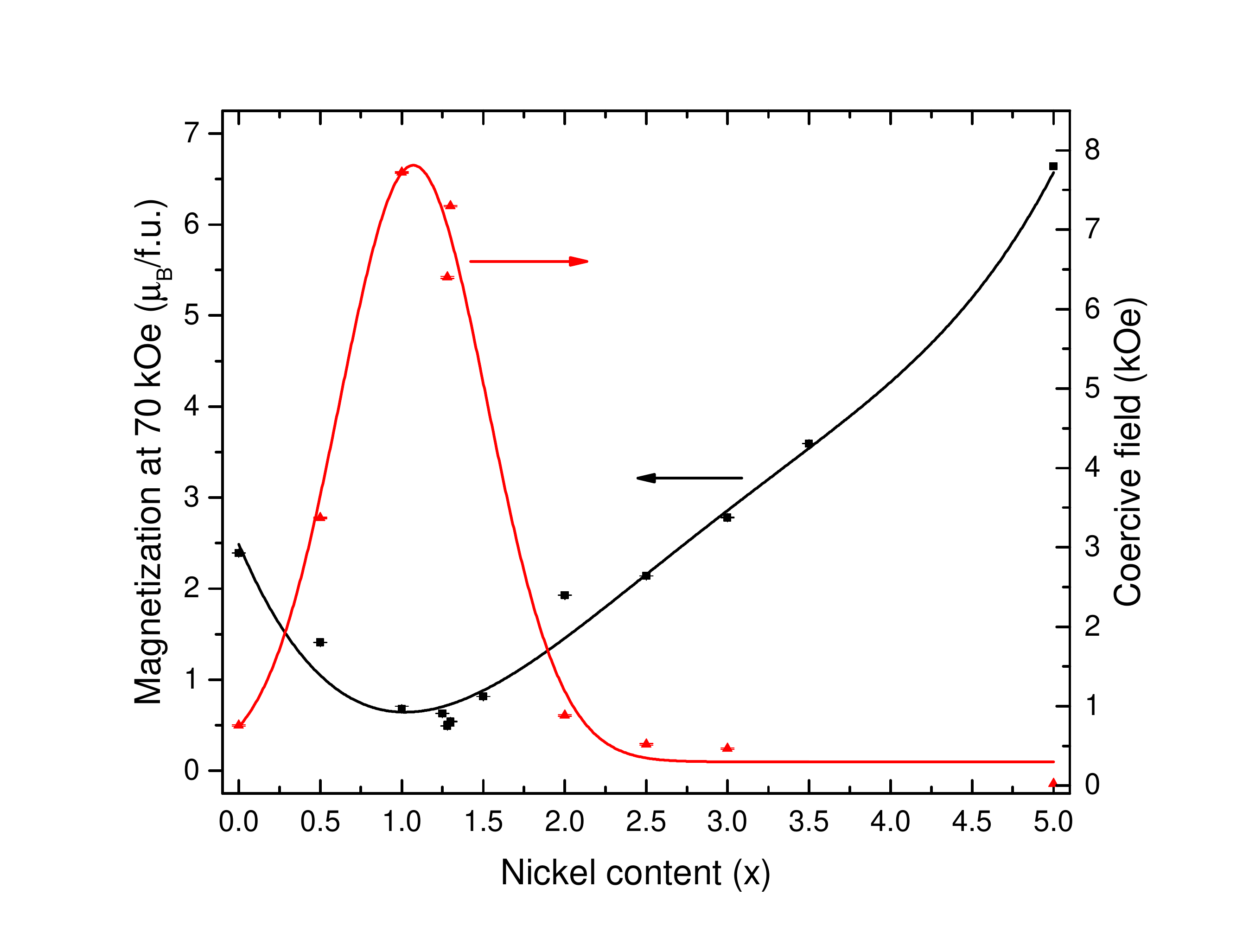}
\caption{\label{fig.MandHc} Magnetization at 70~kOe (black squares) and coercive 
field (red triangles) as a function of Ni content, $x$,  measured
at 10~K (5~K for $x$ = 0, 0.5 and 1).
Errors were estimated from the precision of the 
magnetization measurements.
Lines are a guide to the eye.}
\end{figure}

The magnetization at 70~kOe and the coercive field obtained from the 
$M$ versus $H$ loops on powders/buttons as discussed above are 
shown in Fig.~\ref{fig.MandHc} as a function of composition.
The magnetization has a minimum at $x \sim 1$, which is the 
composition for which the sublattice magnetizations 
cancel maximally in the presence of a 70~kOe applied field at 10~K.
This is consistent with the value predicted in section \ref{sec.intro} 
and with the values obtained via the theoretical approach, 
section \ref{sec.theory}.
The coercive field has a broad peak that corresponds with 
the minimum in magnetization.
This behavior is consistent with that observed for single 
crystals of Cu-doped GdCo$_5$,~\cite{Grechishkin2006, deOliveira2011} 
and other RECo$_{5-x}$Ni$_x$ materials~\cite{Buschow1976, Chuang1982}.
A corresponding peak in magnetocrystalline anisotropy is found 
in the theoretical calculations  discussed in \ref{sec.theory}. 

\section{Theory\label{sec.theory}}

\subsection{Zero temperature magnetization}

\begin{figure}
\includegraphics[width=90mm]{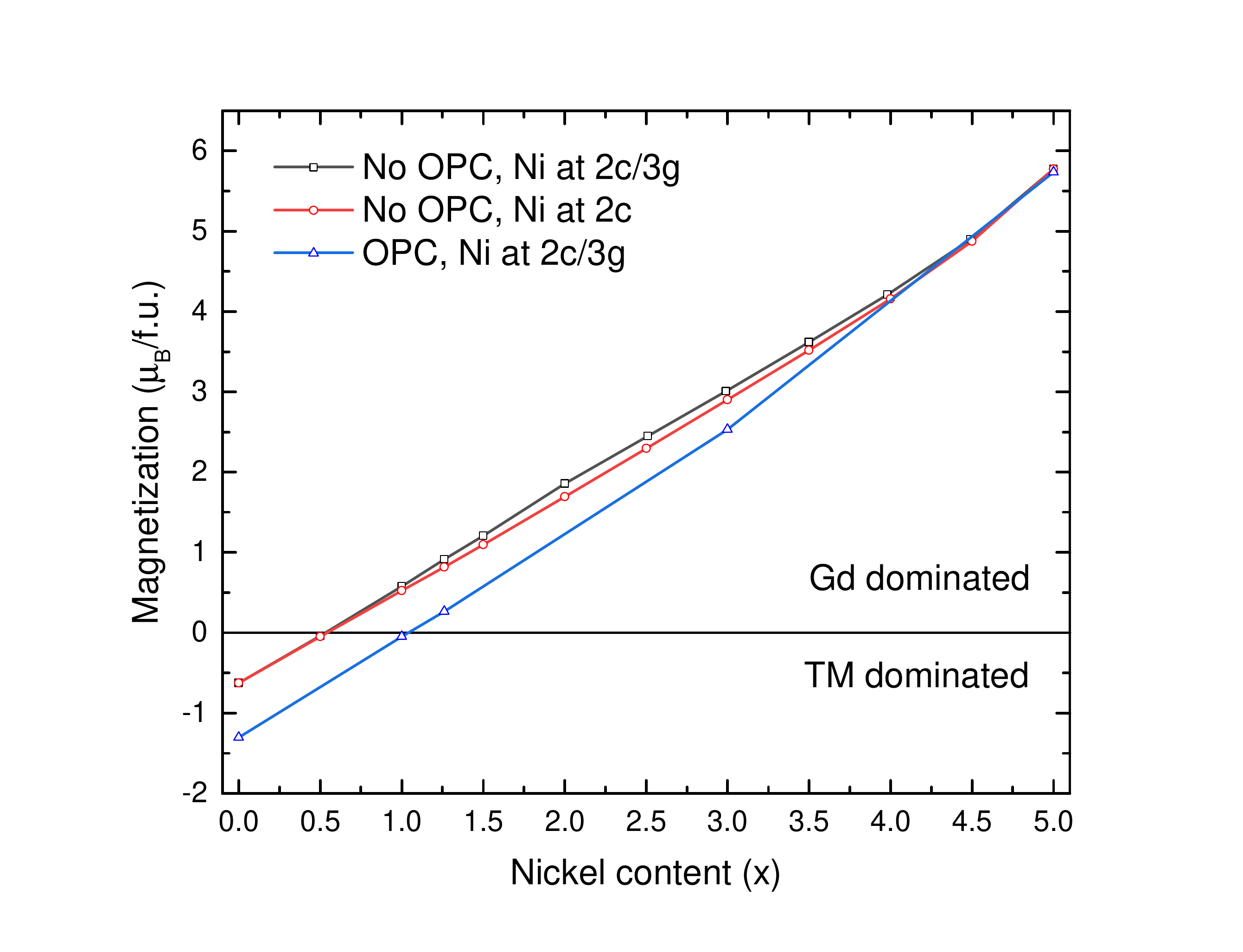}
\caption{\label{fig.Mvx}
Calculated zero-temperature magnetization of GdCo$_{5-x}$Ni$_x$.
A negative value of $M$ (``TM dominated'') implies that the total 
moment points in the same direction as the transition metal 
(Co and Ni) sublattice and opposite to the Gd sublattice, 
and vice versa for a positive value (``Gd dominated'').
The different symbols correspond to calculations: without the 
OPC and with preferential Ni occupation at the $2c$ sites 
(red circles); without the OPC and with equal Ni occupation 
at the $2c$ and $3g$ sites (black squares); with the 
OPC and with equal Ni occupation at the $2c$ and $3g$ 
sites (blue triangles). }
\end{figure}

Figure~\ref{fig.Mvx} shows zero temperature DFT-DLM calculations 
of the magnetization of GdCo$_{5-x}$Ni$_x$ as a function of 
nickel content $x$.
The antiferromagnetic coupling of the RE and TM moments means 
that the total moment is obtained as the difference between 
these two contributions.
For GdCo$_5$, without the orbital polarisation correction a total moment of 0.62$\mu_{\mathrm{B}}$/f.u. is calculated, 
which consists of a contribution from the Co sublattices 
of $(2\times1.65 + 3\times1.61 = 8.13\mu_{\mathrm{B}})$ and 
from the Gd atom of 7.49$\mu_{\mathrm{B}}$.
The main effect of the OPC is to increase the orbital 
moments on each Co atom by $\approx~0.1\mu_{\mathrm{B}}$, 
giving an increased Co contribution to the moment 
of $(2\times1.79 + 3\times 1.73 = 8.77\mu_{\mathrm{B}})$ 
and total moment of 1.30$\mu_{\mathrm{B}}$/f.u. 
The theoretical justification for including the OPC is that it 
approximates the contribution to the exchange-correlation 
energy from the orbital current, which is missing in the local 
spin-density approximation~\cite{Eschrig2005}.
Practically, previous work both on YCo$_5$ and GdCo$_5$ 
found that including the OPC improved the agreement of 
magnetic moments and magnetocrystalline anisotropy with 
experiment~\cite{Steinbeck2001, Patrick2018, Daalderop1996}.

Considering the other limit of GdNi$_5$, the Ni sublattices 
give a much weaker contribution of 
$(2\times0.22 + 3\times0.35 = 1.49\mu_{\mathrm{B}})$ (no OPC). 
The weaker TM magnetism leads to a smaller induced contribution 
to the Gd moment, whose total value is reduced to 7.27$\mu_{\mathrm{B}}$.
The total GdNi$_5$ moment is therefore 5.78$\mu_{\mathrm{B}}$/f.u. 
The OPC has a much smaller effect on the Ni orbital moments 
compared to Co, so that the OPC-calculated total moment is 
reduced only slightly, to 5.73$\mu_{\mathrm{B}}$/f.u.

The absolute value of the moment of GdNi$_5$ exceeds that of GdCo$_5$.
The difference is that, in GdCo$_5$ the total moment points in 
the same direction as the transition metal moments (TM dominated), 
whereas in GdNi$_5$ the total moment points in the same direction as 
the Gd moment (Gd dominated).
In Fig.~\ref{fig.Mvx} the sign convention is adopted that Gd (TM) 
dominated systems have positive (negative) moments.
The gradual addition of Ni weakens the TM contribution, causing
a compositionally-induced transition from TM to Gd dominated magnetism 
with increasing $x$.

At the concentration when the TM and Gd contributions to the magnetization are 
equal, the moments are fully compensated and the total magnetic moment 
of GdCo$_{5-x}$Ni$_x$ is zero.
Since the OPC increases the TM moment whilst leaving the Gd moment 
largely unaffected, this compensation concentration is different 
for calculations with and without the OPC.
With or without the OPC, compensation occurs at $x=1.04$ or $x=0.54$ respectively.
Comparing the calculations with the experimentally-estimated compensation 
concentration of $x\approx1$ also supports the use of the OPC for GdCo$_{5-x}$Ni$_x$.

The calculations discussed above were performed assuming that the Ni atoms 
substitute onto the $2c$ and $3g$ sites with equal probability, and 
are shown as the blue triangles and black squares in Fig.~\ref{fig.Mvx} 
(with and without OPC, respectively).
However, previous calculations found it to be more energetically 
favorable for Ni to substitute at the $2c$ sites~\cite{Patrick2017}.
Neutron diffraction experiments on Ni-doped YCo$_5$ also found this 
preferential $2c$ occupation~\cite{Deportes1976}.
To investigate how this site preference affects the magnetic properties, 
calculations were also performed where the Ni atoms fill the $2c$ sites first, 
with the $3g$ sites only becoming occupied with Ni atoms for $x>2$. 
The moments calculated in this way (without the OPC) are shown as the red 
circles in Fig.~\ref{fig.Mvx}.
The location of the Ni dopants does not have a large effect on the calculated 
moment, yielding a maximum difference of $0.16\mu_{\mathrm{B}}$/f.u. at $x=2$.
The compensation concentration is also unaffected.
However, as discussed below, there is a more pronounced effect on the 
magnetocrystalline anisotropy from site-preferential doping.

\subsection{Temperature dependent magnetization}

\begin{figure}
\includegraphics[width=90mm]{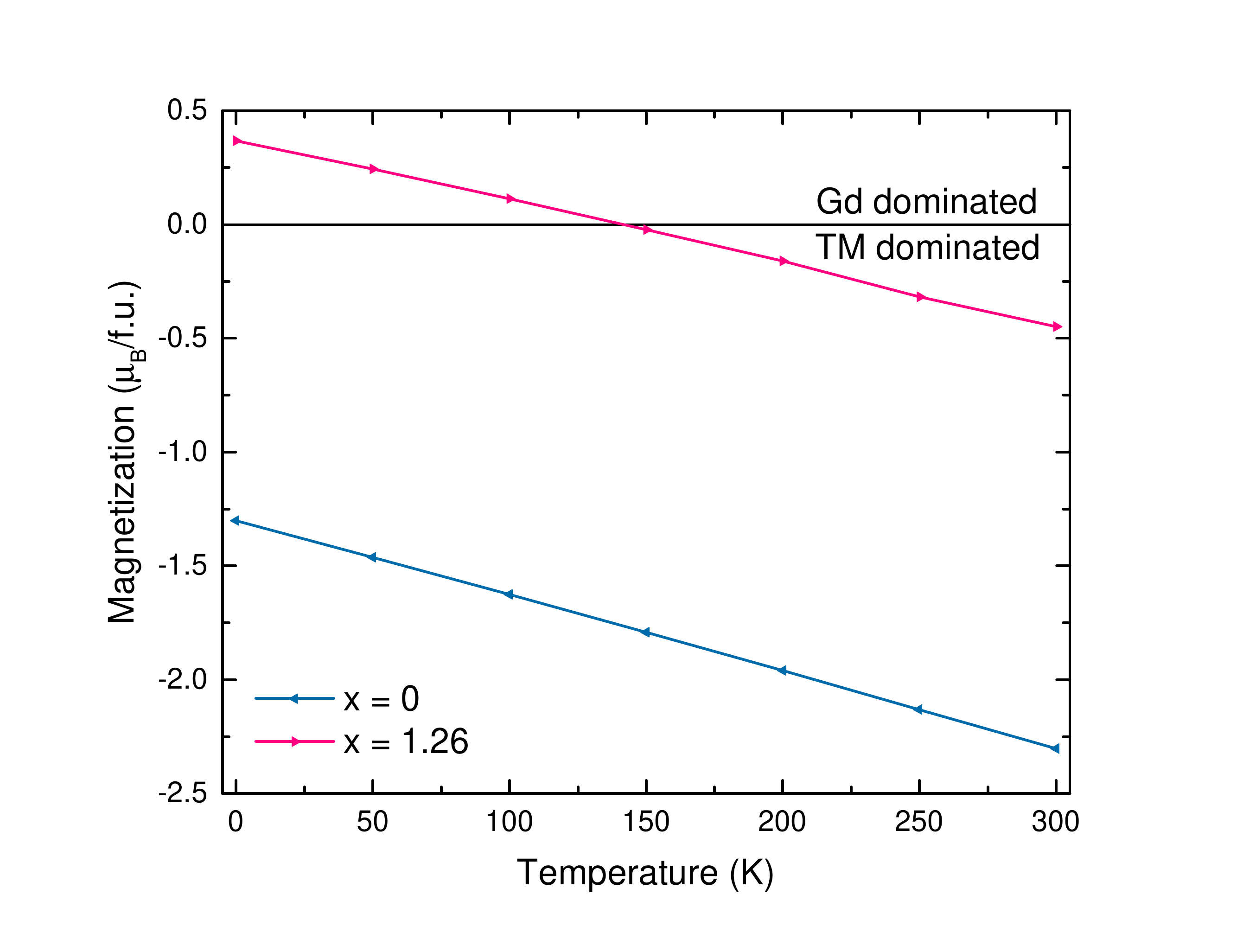}
\caption{\label{fig.MvT}
Calculated magnetization versus temperature for $x~=~0$ (blue triangles)
and $x~=~1.26$ (pink half-filled squares). The sign convention (TM/Gd dominated) is as in Fig.~\ref{fig.Mvx}. }
\end{figure}

We now consider the magnetization at finite temperature, focusing 
on two cases: pristine GdCo$_5$ and GdCo$_{3.74}$Ni$_{1.26}$.
The latter Ni concentration was selected due to the interesting coercivity 
behavior observed experimentally for samples around this composition, 
as shown in Fig.~\ref{fig.MandHc}.
In these calculations the OPC was included, and the Ni dopants occupied the 2$c$ sites only.

Figure~\ref{fig.MvT} shows the DFT-DLM magnetizations calculated for 
the temperature range 0--300~K. As in Fig.~\ref{fig.Mvx}, positive 
values correspond to the Gd moment having a larger moment than 
the TM contribution.
The magnetization of both GdCo$_5$ and GdCo$_{3.74}$Ni$_{1.26}$ becomes 
more negative (TM-dominated) in this temperature range.
The change is effectively linear with temperature, with a difference 
of 1.0$\mu_B$/f.u. for GdCo$_5$ and $0.8\mu_{\mathrm{B}}$/f.u. 
for GdCo$_{3.74}$Ni$_{1.26}$ between 0 and 300~K.

The origin of the change in magnetization is a faster disordering of 
Gd moments compared to the TM as the temperature is 
increased~\cite{Patrick2017}.
This disordering is quantified by the order parameters 
$(m_\mathrm{Gd},m_\mathrm{TM})$, which vary between 1 at 0~K and 
zero at the Curie temperature.
At 300~K, $\left(m_\mathrm{Gd},m_\mathrm{TM}\right) = \left(0.75,0.91\right)$ 
in GdCo$_5$ and (0.75,0.83) in GdCo$_{3.74}$Ni$_{1.26}$.
Therefore in both cases the relative strength of the TM contribution 
compared to Gd has increased with increasing temperature, 
producing a shift towards TM dominated magnetization.

The fact that $m_\mathrm{Gd} = 0.75$ for both cases at 300~K 
shows that the introduction of Ni at the $2c$ sites has not affected 
the rate of Gd disordering, consistent with results obtained 
previously~\cite{Patrick2017}.
However, the presence of Ni does lead to a faster disordering of 
TM moments ($m_\mathrm{TM} = 0.83$ compared to 0.91), which is 
why the change in magnetization between 0--300~K is smaller 
for GdCo$_{3.74}$Ni$_{1.26}$ than GdCo$_5$.
Overall, this faster disordering reduces the Curie temperature, 
which is calculated to be 915~K for GdCo$_5$ and 713~K for 
GdCo$_{3.74}$Ni$_{1.26}$.
These values are consistent with the experiments of Chuang \textit{et al}.~\cite{Chuang1981}, 
who observed a Curie temperature of 1000~K for GdCo$_5$ and 
730~K for GdCo$_{3.75}$Ni$_{1.25}$.

As shown in Fig.~\ref{fig.MvT}, at 140~K GdCo$_{3.74}$Ni$_{1.26}$ 
switches from Gd to TM dominated magnetization.
This temperature, where the antiparallel Gd and TM moments 
cancel each other, is the calculated compensation point 
of this composition, and agrees well with the experimental 
data shown in Fig.~\ref{fig.Tcomp1T}.
In passing, we note that not including the OPC shifts the 
magnetization to a more negative value by 0.55$\mu_{\mathrm{B}}$/f.u. 
at 0~K, and raises the compensation temperature to $\sim$300~K (not shown).

\subsection{Zero temperature magnetocrystalline anisotropy}

\begin{figure}
\includegraphics[width=90mm]{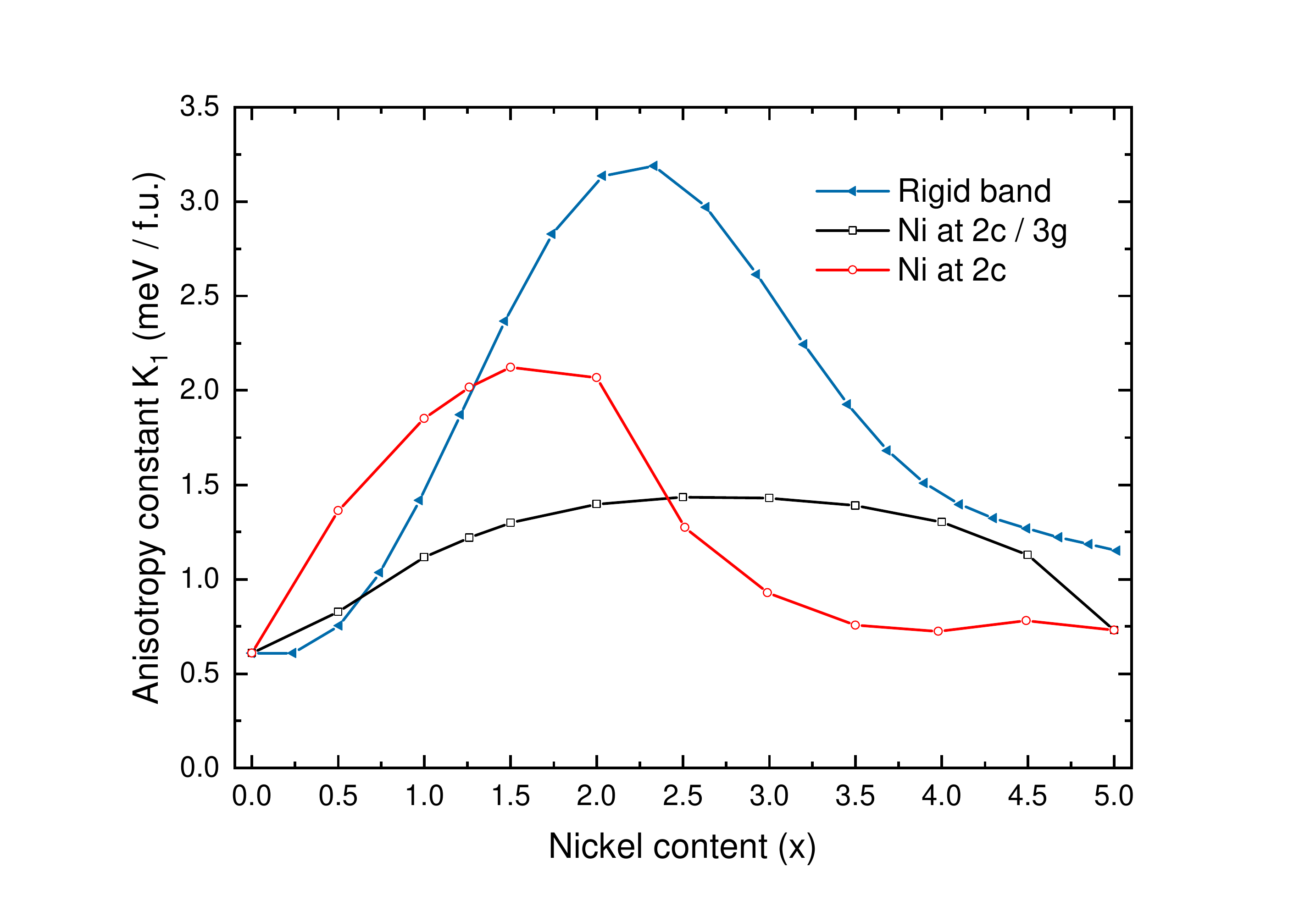}
\caption{\label{fig.Kvx}
Anisotropy energy corresponding to a rigid rotation of the antiparallel TM 
and Gd sublattices, calculated at zero temperature, for GdCo$_{5-x}$Ni$_x$.
The different symbols correspond to preferential Ni substitution at 
the 2$c$ sites (red circles), equal substitution over the 2$c$/3$g$
sites (black squares), or a rigid band calculation on 
pristine $x~=~0$ (blue triangles). }
\end{figure}

We next consider the experimentally-observed variation in coercive 
field with composition (Fig.~\ref{fig.MandHc}).
Arguably the simplest model of coercivity is based on magnetization 
rotation (the Stoner-Wohlfarth [SW] model)~\cite{Chikazumi1997} 
which gives a coercive field of $2K/M$ for a ferromagnet 
of anisotropy $K$ and magnetization $M$.
The same expression is obtained for the nucleation of 
reverse domains within micromagnetic theory~\cite{Brown1957}.
Postponing a discussion of $M$ to Sec.~\ref{sec.twosub}, we first
consider the magnetocrystalline anisotropy of GdCo$_{5-x}$Ni$_x$.
At zero temperature, the angular variation of the free energy was 
calculated, when the Gd and TM moments are held antiparallel to each 
other and rotated from being parallel to perpendicular to the 
crystallographic $c$ axis.
This variation is well described by 
$E_\mathrm{an} (\theta) = K_1 \sin^2\theta + K_2 \sin^4\theta$, 
with $K_2 \ll K_1$. Fig.~\ref{fig.Kvx} shows $K_1$ as a function of Ni composition $x$.
The dominant contribution to this anisotropy energy is the TM sublattice, with
a minor $5d$ contribution from Gd~\cite{Patrick2018}.

As for the zero temperature magnetization in Fig.~\ref{fig.Mvx}, 
both preferentially substituting the Ni at 2$c$ sites 
(circles in Fig.~\ref{fig.Mvx}) and equally distributing the Ni over 
the 2$c$ and 3$g$ sites (squares) was investigated.
In both cases, adding Ni increases $K_1$ compared to pristine 
GdCo$_5$.
Furthermore, both cases show a peak in $K_1$ with Ni content.
For preferential 2$c$ substitution this peak occurs for $x$ between 
1.5--2.0, while for equal 2$c$/3$g$ substitution the peak for $x$ 
is between 2.5--3.0.
The enhanced $K_1$ is much more pronounced for preferential 2$c$ 
substitution, becoming 3.5 times larger compared to 
pristine GdCo$_5$ at $x=1.5$.

In these calculations, the Ni doping has been modeled using the CPA. 
In a simpler rigid-band calculation, the effects of Ni-doping are simulated 
by shifting the Fermi level of pristine GdCo$_5$ so that the integrated 
density of states equals the number of electrons in the Ni-doped system.
The rigid band calculations of $K_1$ are shown as the blue triangles 
in Fig.~\ref{fig.Kvx}.
Here, the enhancement in $K_1$ with $x$ is even greater than 
that found with the CPA.
The rigid band model does not provide a fully consistent picture
of doping, e.g. with the value of $K_1$ 
at $x=5$ not coinciding with $K_1$ calculated for GdNi$_5$.
Nonetheless, the rigid band data emphasizes how, as has 
been previously discussed for YCo$_5$~\cite{Daalderop1996, Steinbeck2001}, 
changing the occupations of the bands located close to the 
Fermi level can have large effects on the anisotropy.

The calculations in Fig.~\ref{fig.Kvx} were performed without the OPC. 
Calculations including the OPC show the same variation with band 
filling, but the values of $K_1$ are strongly enhanced, as observed 
previously for YCo$_5$~\cite{Daalderop1996, Steinbeck2001}.
For instance, for $x=1.26$ with preferential 2$c$ Ni doping, 
values of 2.0~and~6.5~meV/f.u. for $K_1$ without and with 
the OPC, respectively, are found.

\subsection{Zero temperature coercivity}
\label{sec.twosub}

The previous section showed that increasing the Ni content
causes a boost to the anisotropy energy of the transition metal
sublattice.
Assuming Ni substitutes preferentially at 2$c$ sites,
the calculated peak in anisotropy and the experimentally measured peak
in coercivity are located at similar concentrations.
This observation may explain the increased coercivity with Ni doping 
of RECo$_5$ compounds with nonmagnetic REs~\cite{Buschow1976, Chuang1982}.
However, as shown in Fig.~\ref{fig.MandHc}, the maximum in the coercive field for
GdCo$_{5-x}$Ni$_x$
coincides with a minimum in magnetization.
Referring again to the micromagnetic expression for
the coercive field of a ferromagnet of $2K/M$, we note that 
naively setting $M$ to zero 
at finite $K$ should yield a divergent coercive field at the 
compensation point.
This divergence remains even when
Kronm\"uller's prefactor $\alpha$~\cite{Kronmuller1987}
is introduced in order to account for microstructural variation in $K$.
Therefore the boost in coercivity in GdCo$_5$ may simply result from compensation
of the Gd and TM magnetic moments.

However, GdCo$_{5-x}$Ni$_x$ is a ferrimagnet, so it is by no 
means obvious that models based on the rotation of a single 
magnetization vector should apply.
Extending the SW model for a ferrimagnet produces a two-sublattice model, 
which was investigated for positive applied fields in 
Ref.~\citenum{Radwanski1986}.
Crucially, the competition between the external field, the antiparallel 
exchange interaction and the magnetocrystalline anisotropy can lead to 
canting between the Gd and the TM sublattices when a magnetic field is applied.

We recently introduced a method of calculating magnetization versus field curves
including this effect from first principles, which we applied to 
GdCo$_5$ ($x~=~0$) at low~\cite{Patrick2018} and 
high~\cite{PatrickKumar2018} magnetic fields.
In this approach,  DFT-DLM calculations are used to 
parameterize the following expression for the free energy $F_2$.
\begin{eqnarray}
F_2(\theta_\mathrm{Gd},\theta_\mathrm{TM}) &=&  K_{1,\mathrm{TM}} \sin^2\theta_\mathrm{TM}
- \mu_0\mathbf{M}\cdot\mathbf{H} \nonumber \\
&& + K_{2,\mathrm{TM}} \sin^4\theta_\mathrm{TM} + K_{1,\mathrm{Gd}} \sin^2\theta_\mathrm{Gd}  \nonumber \\
&& + S(\theta_\mathrm{TM},\theta_\mathrm{Gd}) + A \ \mathbf{\hat{M}_\mathrm{Gd}} \cdot \mathbf{\hat{M}_\mathrm{TM}}
\label{eq.F2}
\end{eqnarray}
with $\mathbf{M} = \mathbf{M_\mathrm{Gd}} + \mathbf{M_\mathrm{TM}}$.
The first line of Eq.~\ref{eq.F2} resembles the free energy
found in the Stoner-Wohlfarth model.
$\theta_i$ denotes the angle that the magnetization of sublattice $i$ makes with the
$c$ axis, $K_{j,i}$ represents the various anisotropy constants, and $S$ represents the anisotropy
energy due to dipolar interactions.
$A$ quantifies the  exchange interaction, which with a positive
value favors antiferromagnetic alignment of the Gd and TM moments.

For a compensated magnet, $M_\mathrm{Gd} = M_\mathrm{TM}$ 
and in the absence of an external field the magnetic moments are antiparallel.
Naively we might therefore set $\mathbf{M} = 0$ and, from inspection of 
equation~\ref{eq.F2}, argue that the external field can have no effect on 
the free energy or magnetization, corresponding  to infinite coercivity.
However, $\mathbf{M} = 0$ is only true as long as the moments remain
antiparallel.
If the antiparallel alignment breaks, the magnetic sublattices couple
individually to the external field.
For instance, in the limit of extremely strong external fields both sublattices
align to the field, giving a resultant magnetization of 
$M_\mathrm{Gd} + M_\mathrm{TM}$.

We note that this model provides an explanation for the magnetization measurements of
the powder in a 10~kOe field as a function of temperature (Fig.~\ref{fig.MT}).
On free-to-rotate samples, the critical field required to trigger the
transition from antiparallel to canted moments essentially scales as
$|M_\mathrm{Gd} - M_\mathrm{TM}|$~\cite{PatrickKumar2018,Isnard2012}.
Therefore, as one approaches the compensation point in the (free-to-rotate) powder,
the antiparallel alignment can be broken with a small field, and a nonzero
magnetic moment measured.

\begin{figure}
\includegraphics[width=90mm]{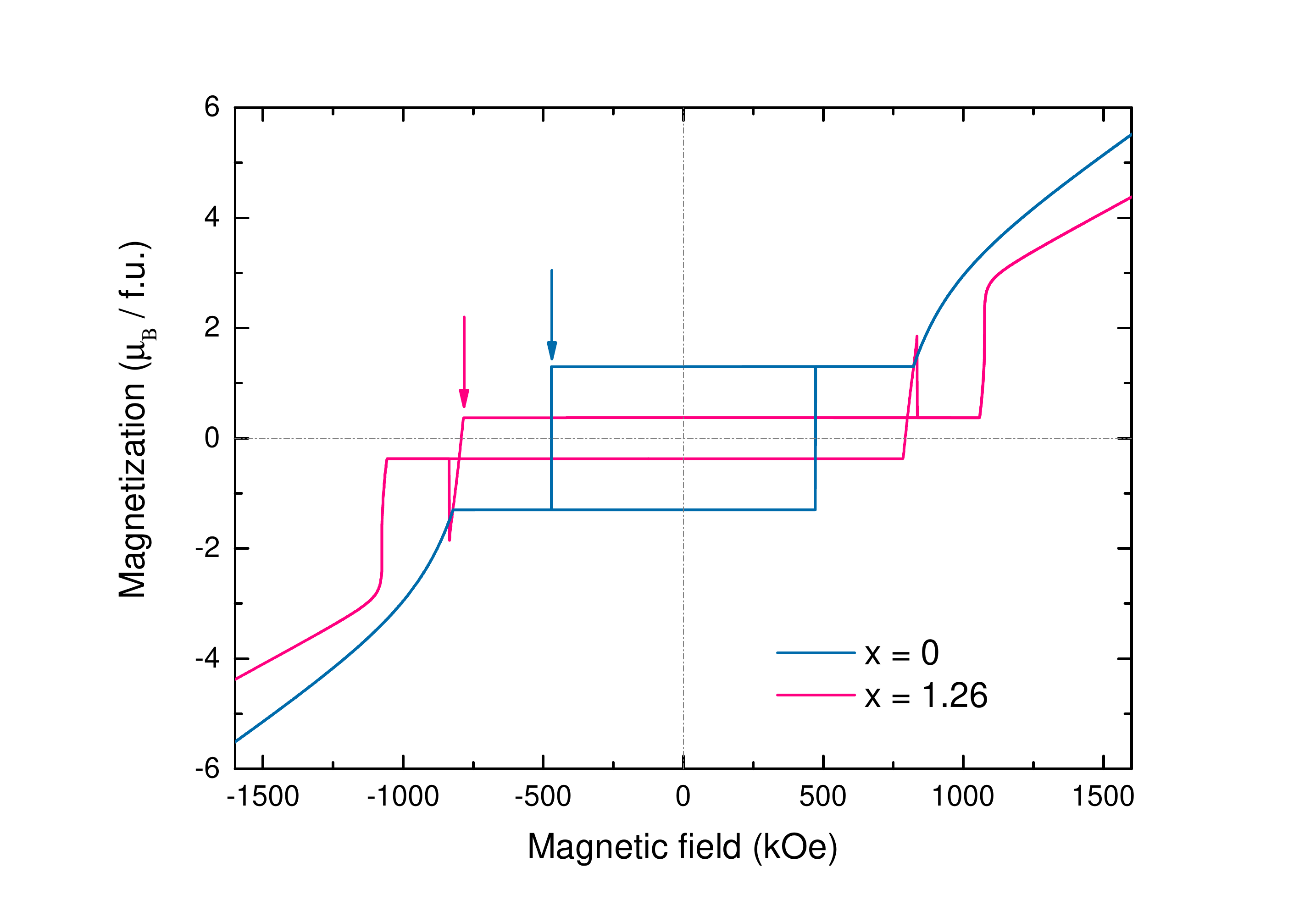}
\caption{\label{fig.5quad}
Magnetization versus field curves calculated at zero temperature for
$x~=~0$ (blue) and $x~=~1.26$ (pink).
The arrows label the coercive fields at which the magnetization 
switches from positive to negative values during the 
downward field sweep (second quadrant).
}
\end{figure}
Now, considering fixed samples,
in Fig.~\ref{fig.5quad}, the results of minimizing $F_2$ 
along the full multi-quadrant magnetization curve are shown, 
sweeping the field along the sequence
$0\rightarrow H_\mathrm{max} \rightarrow -H_\mathrm{max} \rightarrow H_\mathrm{max}$,
for GdCo$_5$ and GdCo$_{3.74}$Ni$_{1.26}$.
Here the OPC is included, preferential Ni doping at the 2$c$ 
sites is assumed (cf.\ Fig.~\ref{fig.MvT}), and the calculations 
performed at zero temperature.
The field is applied along the crystallographic $c$ axis.
The size of the field is not intended to match experimental
results, as described below, but the relative changes between 
different compositions can be extracted.

Focusing first on GdCo$_5$ (blue line) for $\left|H\right| < 820$~kOe,
the boxlike curve resembles that of a SW ferromagnet.
At 820~kOe, there is a transition from the rigid antiparallel alignment 
of Gd and Co moments to a canted configuration, with the energy gain 
of the Gd moments aligning with the magnetic field competing with 
the exchange and anisotropy terms.
This transition is reversible, such that there is no hysteresis in 
the first quadrant.
In the second quadrant, at $H = -473$~kOe 
(blue arrow in Fig.~\ref{fig.5quad}) there is a discontinuous jump 
in the magnetization corresponding to a simultaneous 180$^\circ$ 
rotation of the Gd and Co moments.
This jump is irreversible, so returning the field to zero now gives 
a negative magnetization, with the majority of the Co moments 
now pointing opposite to the field until the symmetric jump at 
$H = 473$~kOe occurs.

GdCo$_{3.74}$Ni$_{1.26}$ ($x~=~1.26$, pink line) shows broadly the same 
behavior, but the nature of the transitions themselves are 
slightly different.
At the high-field transition from antiparallel to canted moments 
at $\left|\mu_0H\right| = 1060$~kOe, the moments rotate rapidly 
with field such that there is a very sudden, but reversible, 
increase in magnetization.
However, the demagnetizing curve in the second quadrant shows 
a new feature, which is a continuous and reversible decrease of 
magnetization in the region -835~kOe $< \mu_0H <$ -785~kOe.
The magnetization passes through zero at -792~kOe, and becomes 
increasingly negative, exceeding its zero-field magnitude at -800~kOe.
This new feature is a result of the system getting trapped in a 
metastable energy minimum corresponding to canted Gd and TM moments.
For $\left|H\right|> 835$~kOe this minimum disappears, and 
the system undergoes an irreversible transition back to antiparallel moments.

Equating the coercive fields with the magnitudes of the applied 
fields which produce zero magnetization in the second quadrant, 
we extract values of 473~kOe for $x~=~0$ and 792~kOe for $x~=~1.26$.
For now ignoring the fact that these numbers are huge compared 
to experiment, in terms of relative magnitudes an increase in 
coercivity by a factor of 1.7 is observed at a Ni doping 
of $x=1.26$.
We note that this increase is relatively modest compared to a 
naive prediction based on assuming that the coercivity was 
proportional to $K_1/M$; since $K_1$ and $M$ increase/decrease 
by a factor of 3 respectively, we might have expected a 
coercivity enhancement by a factor of 9.

The calculations in Fig.~\ref{fig.5quad} are illustrative, but 
cannot be considered a realistic picture of macroscopic magnetization 
reversal.
In reality, the nucleation of reverse domains, e.g.\ at the edge 
of the sample, will facilitate magnetization reversal at far 
lower fields than found here~\cite{Kumar1988, Fischbacher2017}.
The coercivity will then depend on how the domain walls propagate 
through the sample, which is likely to be affected by the 
presence of the secondary phase~\cite{SepehriAmin2017}.
We also note that the peak in coercivity observed experimentally 
here was found for polycrystalline samples.
However, it is interesting that the small single crystals of 
Cu-doped GdCo$_5$ reported in Ref.~\citenum{Grechishkin2006} 
do show box-like demagnetization curves such as the calculated
ones shown in Fig.~\ref{fig.5quad}.

\section{Summary \& Conclusions \label{summary}}

Polycrystalline samples of GdCo$_{5-x}$Ni$_x$ for $x=0$ to 5 have 
been synthesized using an arc furnace. 
The predominant formation  of a  single phase was confirmed by comparing powder 
x-ray diffraction patterns to the pattern measured for pure GdCo$_5$.
Optical and SEM imaging showed small ($<$7\%) amounts of a 2:7 phase in 
the 1:5 matrix for $x \leq 2.5$, and no 2:7 phase at higher concentrations. 
No evidence was found to say with confidence that the annealing improves 
(or indeed, affects at all) the microstructure and phase purity of the samples.

The magnetization of the samples measured at 70~kOe and 10~K initially decreases as the
nickel content is increased.
At a composition of $x \approx 1$, the (absolute) magnetization reaches a minimum
and then increases with further Ni addition.
This behavior is due to the Ni weakening the magnetization of the transition
metal sublattice, such that at low temperature, for $x < 1$ the net magnetization
points along the direction of the transition metal moments, while for $x > 1$
the net magnetization points along the direction of the Gd moments.
Zero temperature DFT-DLM calculations find the compensation composition, i.e.\
the point at which the transition metal and Gd sublattice magnetizations cancel each
other to be $x~=~1.04$, in good agreement with the experimentally-observed
minimum.
The calculations found the magnetization to be rather insensitive to the 
location of the Ni dopants, which can occupy either 2$c$ or 3$g$ crystallographic 
sites. 

For a Ni content of $1 \leq x \leq 3$, compensation temperatures in the range 10--360~K
were observed.
The compensation temperatures increase with increasing Ni doping, and occur due to
the faster disordering of the Gd moments compared to the transition metal.
Finite temperature DFT-DLM calculations on pristine GdCo$_5$ and GdCo$_{3.74}$Ni$_{1.26}$
demonstrate this behavior explicitly, finding a compensation temperature of 140~K for
the latter compound which corresponds well to the experimental measurements shown
in Fig.~\ref{fig.Tcomp1T}.

The coercivity of polycrystalline buttons measured below 10~K 
is found to have a maximum value at a composition $x \approx 1$, 
coinciding with the minimum in magnetization. 
One might argue that such behavior is consistent with the Stoner-Wohlfarth model
and micromagnetics, where the coercive field is inversely proportional to the magnetization
($H_c=2K/M$).
However, such a picture is based on the rotation of a single magnetic sublattice and
neglects the possibility that magnetization reversal might proceed via a canted arrangement
of Gd and transition metal moments.
Magnetization versus field loops calculated allowing for such canting
do show an increase in coercivity from $x=0$ to $x=1.26$, but not by as 
great an amount as predicted by the Stoner-Wohlfarth
model given the reduction in $M$.
Apart from the reduction in magnetization, the DFT-DLM calculations also found 
an increase in the magnetocrystalline anisotropy of the transition-metal 
sublattice with Ni doping.
Indeed, assuming preferential substitution at the 2$c$ sites 
the peak in anisotropy was
found for a concentration of $x=1.5$,
reasonably close to the experimentally observed
coercivity maximum.
Such an explanation for increased coercivity, independent of phenomena related to 
compensation, would be consistent with measurements on doped RECo$_5$ compounds
with nonmagnetic RE, which also undergo peaks in coercivity despite having no
compensation points.

However, apart from these intrinsic factors, it should also be noted that the peak
in coercivity also coincides with the largest amount of secondary 
2:7 phase (Table~\ref{tab.secondphase}).
Although the amount of secondary phase we observe is small in terms of measuring
intrinsic quantities, interfaces between the 1:5 and 2:7 phases could 
inhibit the motion of domain walls through the sample, increasing 
the coercivity.
Being able to better control the formation of the 2:7 
phase would allow the magnitude of this extrinsic effect to be tested.

The current study emphasizes the complementary roles played by experiments and theory.
On one hand, the experiments provide valuable input for developing the calculations,
particularly in terms of validating the methodology.
On the other hand, the calculations provide microscopic insight into macroscropic
measurements.
Here we have shown that quantitative comparisons are possible between intrinsic
quantities such as magnetizations and compensation temperatures.

However, whilst the calculations can give hints about extrinsic quantities such as
the coercivity, in reality a multiscale approach capable of describing e.g.\ microstructure
and long range demagnetizing fields is required.
Nonetheless, the first-principles calculations (as validated by experimental measurements
of intrinsic quantities) can still play a fundamental role by providing the 
microscopic parameters required as input for such simulations.

\section*{Acknowledgements}
This work forms part of the PRETAMAG project, funded by the UK Engineering 
and Physical Sciences Research Council (EPSRC) Grant No. EP/M028941/1.
Crystal growth work at Warwick was also supported by EPSRC Grant No. EP/M028771/1.
We thank the Warwick Research Technology Platform (X-ray Diffraction and Electron Microscopy) 
for their assistance.
We acknowledge the Warwick Manufacturing Group and Buehler for assistance with 
slide production.
We thank E. Mendive-Tapia and G. Marchant for useful discussions.


%
\end{document}